%% file: 0_main.tex
\documentclass[10pt,conference]{IEEEtran}
\input{macro}

\def\BibTeX{{\rm B\kern-.05em{\sc i\kern-.025em b}\kern-.08em
    T\kern-.1667em\lower.7ex\hbox{E}\kern-.125emX}}

\makeatletter
\newcommand{\linebreakand}{%
  \end{@IEEEauthorhalign}
  \hfill\mbox{}\par
  \mbox{}\hfill\begin{@IEEEauthorhalign}
}
\makeatother

\begin{document}

\title{Automatically Enhancing the Quality of Android App Bug Reports}

\author{
\IEEEauthorblockN{Antu Saha}
\IEEEauthorblockA{
\textit{William \& Mary}\\
Williamsburg, Virginia, USA \\
\href{mailto:}{asaha02@wm.edu}}

\and
\IEEEauthorblockN{Atish Kumar Dipongkor}
\IEEEauthorblockA{
\textit{University of Central Florida}\\
Orlando, Florida, USA \\
\href{mailto:}{atish.kumardipongkor@ucf.edu}}

\and
\IEEEauthorblockN{Sam Bennett}
\IEEEauthorblockA{
\textit{William \& Mary}\\
Williamsburg, Virginia, USA \\
\href{mailto:}{srbennett01@wm.edu}}

\linebreakand
\IEEEauthorblockN{Kevin Moran}
\IEEEauthorblockA{
\textit{University of Central Florida}\\
Orlando, Florida, USA \\
\href{mailto:}{kpmoran@ucf.edu}}

\and
\IEEEauthorblockN{Andrian Marcus}
\IEEEauthorblockA{
\textit{George Mason University}\\
Fairfax, Virginia, USA \\
\href{mailto:}{amarcus7@gmu.edu}}

\and
\IEEEauthorblockN{Oscar Chaparro}
\IEEEauthorblockA{
\textit{William \& Mary}\\
Williamsburg, Virginia, USA \\
\href{mailto:}{oscarch@wm.edu}}
}

\maketitle

\begin{abstract}

Most defects in mobile applications are visually observable on the device screen. Since automated mechanisms for detecting and reporting such defects are often unavailable, users, testers, and developers must manually submit bug reports. However, these reports are frequently incomplete, ambiguous, or inaccurate, often lacking the information needed to understand, reproduce, and diagnose defects. 

We formulate automatic bug report enhancement as the problem of connecting user-written bug reports with application execution. We present \tool, an LLM-powered approach that links bug report information with app-specific UI execution information to infer and generate accurate, complete, and correct Observed Behavior (OB), Expected Behavior (EB), and Steps to Reproduce (S2Rs). 
To support \tool's design and evaluation, we develop a bug report quality model and use it to identify the most effective context for each component. We evaluate \tool on \rev{\TotalTestBugs{}} bug reports from \rev{\TotalUniqueApps{}} Android applications with manually constructed ground truth. Our results show that \tool generates higher-quality bug report components than the original reports and three LLM-based baselines, improving S2R quality by \rev{44.2}\%--\rev{84.7}\%\rev{, and} OB/EB quality \rev{by up to 35.2\%.}
\rev{We further show that \tool-generated reports help an LLM-based agent reproduce Android bugs, improving the reproduction rate by 28.1\%--57.7\% compared to the original reports and the baselines.}
\looseness=-1

\end{abstract}

\IEEEpeerreviewmaketitle

\input{1_intro}

\input{2_background}

\input{3_approach}

\input{4_prompt_dev}

\input{5_evaluation}

\input{7_threats}

\input{8_related_work}

\input{9_conclusions}

\balance
\bibliographystyle{IEEEtran}
\bibliography{references}

\end{document}

%% file: macro.tex
\usepackage[utf8]{inputenc}
\usepackage{url}
\usepackage{cite}   %
\usepackage{colortbl}

\usepackage{balance}
\usepackage{xspace}
\usepackage{graphicx}
\usepackage{array} 
\usepackage{verbatim}
\usepackage{rotating}
\usepackage{bold-extra}
\usepackage{multirow}
\usepackage{listings}
\usepackage{boxedminipage}
\usepackage{soul}
\usepackage{enumitem}
\usepackage[normalem]{ulem}
\setlist{nolistsep,leftmargin=.5cm}
\usepackage{booktabs}
\usepackage{tabularx}
\usepackage{svg}
\usepackage{calc}
\usepackage{float}
\usepackage{multirow}
\usepackage[normalem]{ulem}
\useunder{\uline}{\ul}{}
\usepackage[most]{tcolorbox}
\definecolor{MidnightBlue}{HTML}{006895}
\definecolor{BoxesBlue}{HTML}{DEECFF}
\definecolor{BoxesYellow}{HTML}{FFF2CC}
\definecolor{StateGreen}{HTML}{91C788}
\definecolor{StateRed}{HTML}{FF8080}
\definecolor{ArrowGreen}{HTML}{61B15A}
\definecolor{ArrowViolet}{HTML}{BA94D1}
\usepackage{caption}
\usepackage{microtype} 
\usepackage{amsmath,amsfonts}
\usepackage{algorithmic}
\usepackage{textcomp}
\usepackage{xcolor}
\usepackage{hyperref}
\usepackage{ifthen}
\usepackage{wrapfig}
\usepackage{subcaption}
\usepackage{cleveref}

\usepackage{threeparttable}
\usepackage{makecell}

\newboolean{showcomments}
\setboolean{showcomments}{true}

\ifthenelse{\boolean{showcomments}}
{\newcommand{\nb}[2]{
		\fbox{\bfseries\sffamily\scriptsize#1}
		{\sf\small$\blacktriangleright$\textit{#2}$\blacktriangleleft$}
	}
}
{\newcommand{\nb}[2]{}
}

\usepackage{amssymb}

\newcommand{\ie}{\textit{i.e.},\xspace}
\newcommand{\eg}{\textit{e.g.},\xspace}
\newcommand{\etc}{\textit{etc.}\xspace}
\newcommand{\etal}{\textit{et al.}\xspace}

\newcommand{\tool}{{\textsc{BugScribe}}\xspace}

\newcommand{\bsgpt}{{\textsc{BSgpt}}\xspace}

\newcommand{\bscld}{{\textsc{BSclaude}}\xspace}

\newcommand{\ap}{\tool}

\newcommand\rev[1]{{\color{black}{#1}}}

\newcommand{\gpt}{{\textsc{GPT-5.4}}\xspace}
\newcommand{\claude}{{\textsc{ClaudeOpus-4.6}}\xspace}
\newcommand{\gptfouro}{{\textsc{GPT-4o}}\xspace}
\newcommand{\opusjudge}{{\textsc{ClaudeOpus-5.5}}\xspace}
\newcommand{\gptjudge}{{\textsc{GPT-5.6-Sol}}\xspace}
\newcommand{\qwenjudge}{{\textsc{Qwen3.6-35B-A3B}}\xspace}
\newcommand{\CrashScope}{{\sc CrashScope}\xspace}
\newcommand{\astro}{{\sc AstroBR}\xspace}
\newcommand{\chat}{{\sc ChatBR}\xspace}

\newcommand{\uts}{{\sc U2Sbr}\xspace}

\newcommand{\gptnoinfo}{{\sc GPTno-info}\xspace}
\newcommand{\cldnoinfo}{{\sc CLAUDEno-info}\xspace}

\newcommand{\codex}{{\sc Codex}\xspace}
\newcommand{\cldc}{{\sc ClaudeCode}\xspace}

\newcommand{\mobilerun}{{\sc Mobilerun}\xspace}

\newcommand{\ebug}{{\sc EBug}\xspace}
\newcommand{\burt}{{\sc Burt}\xspace}

\newcommand{\recdroid}{{\sc ReCDroid}\xspace}

\newcommand{\adbgpt}{{\sc AdbGPT}\xspace}

\newcommand{\EulerC}{{\sc Euler}\xspace}
\newcommand{\Fusion}{{\sc Fusion}\xspace}

\usepackage{tikz}

\definecolor{bug_red}{rgb}{.84,.23,.29}

\definecolor{info-needed-color}{rgb}{1,.8,.12}

\definecolor{lightblue}{rgb}{ .753,  .902,  .961}

\definecolor{lightgreen}{rgb}{.596, 1, .596}

\definecolor{lightyellow}{rgb}{1,0.94902,0.8}

\newcommand{\NIfone}{\rev{72.0}}

\newcommand{\INTms}{13.7}

\newcommand{\INTfone}{\rev{87.3}}

\newcommand{\ISes}{16.0}
\newcommand{\ISms}{12.3}

\newcommand{\ISfone}{\rev{87.2}}

\newcommand{\ISBes}{13.7}

\newcommand{\ISBprec}{\rev{87.6}}
\newcommand{\ISBrec}{\rev{88.7}}
\newcommand{\ISBfone}{\rev{88.2}}

\newcommand{\ISBfoneSTD}{0.01}

\newcommand{\rnum}[1]{\textcolor{black}{#1}}

\newcommand{\ConfigNoInfo}{\textcolor{black}{\textit{No Info.}}}
\newcommand{\ConfigInt}{\textcolor{black}{\textit{Interac.}}}
\newcommand{\ConfigIntScr}{\textcolor{black}{\textit{Interac.+Scrn.}}}
\newcommand{\ConfigIntScrBug}{\textcolor{black}{\textit{Interac.+Scrn.+BuggyScrn.}}}

\newcommand{\ConfigBuggyScr}{\textcolor{black}{\textit{Buggy Scrn.}}}

\newcommand{\ConfigStoRBuggyScr}{\textcolor{black}{\textit{S2Rs+BuggyScrn.}}}

\newcommand{\ConfigStoRAllScr}{\textcolor{black}{\textit{S2Rs+BuggyScrn.+Scrns.}}}

\newcommand{\SumNCor}{30.3}

\newcommand{\SumBCor}{32.3}

\newcommand{\SumSSCor}{34.3}
\newcommand{\SumSSInc}{4.3}

\newcommand{\devOBagr}{\rnum{87.5\%}}
\newcommand{\devOBck}{\rev{\rnum{0.60}}}
\newcommand{\devOBka}{\rnum{0.94}}

\newcommand{\devSTRagr}{\rnum{99.5\%}}
\newcommand{\devSTRck}{\rnum{0.98}}
\newcommand{\devSTRka}{\rnum{0.99}}

\newcommand{\OrigBRcs}{\rev{193}}

\newcommand{\OrigBRfone}{\rev{48.5}}

\newcommand{\UtwoScs}{243}
\newcommand{\UtwoSes}{75}
\newcommand{\UtwoSms}{356}

\newcommand{\UtwoSfone}{\rev{53.0}}

\newcommand{\GPTNIcs}{345}

\newcommand{\GPTNIfone}{\rev{61.2}}

\newcommand{\ClaudeNIcs}{312}

\newcommand{\ClaudeNIfone}{\rev{59.7}}

\newcommand{\BSGPTcs}{532}
\newcommand{\BSGPTes}{57}
\newcommand{\BSGPTms}{67}
\newcommand{\BSGPTprec}{\rev{90.3}}

\newcommand{\BSGPTfone}{\rev{89.6}}

\newcommand{\BSClaudecs}{538}

\newcommand{\BSClauderec}{\rev{89.8}}
\newcommand{\BSClaudefone}{\rev{88.3}}

\newcommand{\TstTSROrigCor}{12}

\newcommand{\TstTGIbCor}{42}         
\newcommand{\TstTSRBcor}{30}

\newcommand{\TstTSRCCCOr}{35}   \newcommand{\TstTSRCCCmis}{2}

\newcommand{\TotalTestBugs}{\rnum{48}}
\newcommand{\TotalOutputBugs}{\rnum{18}}
\newcommand{\TotalCosmeticBugs}{\rnum{12}}
\newcommand{\TotalCrashBugs}{\rnum{12}}
\newcommand{\TotalNavigationBugs}{\rnum{6}}

\newcommand{\tstAvgScrn}{\rnum{43.4}}
\newcommand{\tstMinScrn}{\rnum{5}}
\newcommand{\tstMaxScrn}{\rnum{105}}
\newcommand{\tstMedScrn}{\rnum{27.5}}

\newcommand{\tstAvgInter}{\rnum{117.5}}
\newcommand{\tstMinInter}{\rnum{10}}
\newcommand{\tstMaxInter}{\rnum{282}}
\newcommand{\tstMedInter}{\rnum{92.5}}

\newcommand{\TotalStRTestSet}{\rnum{599}}
\newcommand{\AVGStRTestSet}{\rnum{12.5}}

\newcommand{\TotalUniqueApps}{\rnum{26}}

%% file: 1_intro.tex
\section{Introduction}
\label{sec:intro}

Most defects in mobile applications are visually observable on the device screen~\cite{johnson2022empirical}. Automated mechanisms for detecting and reporting such bugs are often unavailable, especially when the bugs do not produce explicit signals such as crashes. As such, users, testers, and developers must manually submit bug reports (BRs) during app use or testing~\cite{song2022toward}, and developers rely on these reports to understand, diagnose, reproduce, and fix the defects~\cite{Bettenburg2008GoodBR}. A high-quality bug report describes at least three essential components~\cite{Zimmermann2010,Bettenburg2008GoodBR}. The \textbf{Observed Behavior (OB)} describes the faulty behavior, the \textbf{Expected Behavior (EB)} describes the intended behavior, and the \textbf{Steps to Reproduce (S2Rs)} describe the actions needed to reproduce the defect.

Unfortunately, bug reports are often incomplete, ambiguous, or inaccurate, and sometimes omit the EB or S2Rs entirely~\cite{Chaparro2017}. Low-quality reports hinder developers' understanding of defects~\cite{Chaparro2019}, delay resolution~\cite{Guo2010,Zimmermann2012}, and may lead to reopened or unresolvable issues~\cite{Zimmermann2012,Guo2010}. This problem is particularly challenging for mobile apps, whose defects are often GUI-centric and involve multiple screens, interactions, and execution states. End users find these difficult to describe precisely, since they are typically unfamiliar with app internals and with the information developers need. The core challenge in \textit{enhancing} these reports is therefore not merely improving the bug description. It is \textit{recovering the connections between the reported information and the app's underlying UI execution}.
\looseness=-1

Prior work has proposed techniques to assess bug report quality~\cite{mahmud2025combining,Chaparro2019}, identify missing information~\cite{song2020bee,Bo2024}, provide feedback to reporters~\cite{mahmud2025combining,Chaparro2019}, support interactive report construction~\cite{song2022toward,song2022burt,Fazzini:TSE22,Moran2015}, and restructure bug reports~\cite{acharya2025can}. However, these approaches either leave users responsible for correcting the report or generate content primarily from the original report text. Hence, they cannot ensure that the generated information reflects the application's behavior, and may produce incorrect or incomplete descriptions, such as S2Rs that do not correspond to valid GUI interactions.

Our key insight is that high-quality bug report generation should be formulated as a problem of \textit{connecting bug descriptions with the application's UI execution information}. The original report contains information that identifies the relevant screens, GUI interactions, execution paths, and buggy screen, but these relationships are implicit. By combining LLM reasoning with app-specific execution information, these connections can be inferred and translated into detailed, complete, and accurate OB, EB, and S2R descriptions. However, we posit that different bug report components require different execution information, which motivates a component-specific grounding strategy that provides the most relevant context for each component.
\looseness=-1

In this paper, we present \tool, an LLM-powered approach that automatically enhances user-written Android bug reports. \tool constructs a graph-based execution model of app screens and GUI interactions enriched with UI metadata, and identifies the OB, EB, and S2R sentences in the original report. It then links this information to relevant elements of the execution model and generates each report component with a component-specific prompt. \rev{\tool's main novelty is this component-specific strategy, derived from our empirical analysis, which selectively combines bug report text with app execution information according to the requirements of each bug report component.}
\looseness=-1

To design \tool, we used a data-driven methodology to identify which combinations of app-specific information are most effective for generating each component. The analysis uses a bug report quality model that we developed by extending an existing S2R quality model~\cite{Chaparro2019} to include the OB and EB. Using this model and manually curated ground truth for 10 bug reports from nine Android apps, we evaluated eight context configurations. The results show that S2R generation depends primarily on GUI interactions and the buggy screen, whereas OB and EB generation benefits most from the generated S2Rs, the buggy screen description, and screen-level context.

We further evaluated \ap on \TotalTestBugs~bug reports from \TotalUniqueApps~Android apps, for which we manually constructed ground truth OB, EB, and S2Rs following our quality model. Compared to the original reports and three LLM-based baselines that do not use app-specific information, \ap generates more complete, detailed, and accurate S2Rs, with a relative improvement of \rev{44.2}\%\rev{--}\rev{84.7}\%, and improves OB and EB quality \rev{up to 35.2\%.}
\rev{To assess the usefulness of \ap's reports, we used an LLM-based agent, \mobilerun~\cite{mobilerun}, to reproduce 40 bugs with each report version.
With \ap's reports, the agent reproduces 68.3\% of the bugs, compared to 43.3\%--53.3\% with the original and baseline reports, obtains more consistent outcomes across runs, and reproduces four bugs not reproduced \rev{with the original or baseline reports}.}
\looseness=-1

In summary, this paper makes the following contributions:
\begin{itemize}
\item \ap, an LLM-based approach that links bug report information to execution information to generate accurate, complete, and correct OB, EB, and S2R descriptions.
\item A bug report quality model for OB, EB, and S2R descriptions, and a data-driven methodology that uses it to identify the app information elements \ap needs.
\item \rev{An empirical evaluation showing that \ap improves original bug reports, outperforms LLM-based baselines, and helps an LLM-based agent reproduce more bugs.}
\item \rev{A replication package with a manually curated dataset of 58 bug reports, ground truth OB/EB/S2Rs, code, prompts, documentation, experimental infrastructure, and the bug reproduction runs with their evaluation~\cite{package}.}
\end{itemize}

%% file: 2_background.tex
\section{Bug Report Quality Model}
\label{sec:example_q_model}

Our quality model defines quality categories for the three main bug report components: the OB, EB, and S2Rs. A high-quality bug report must contain complete, detailed, and accurate descriptions of these components.
The model is \rev{specific} for reports of GUI-based applications, in particular Android apps.

\textbf{S2R Quality Model.} The S2Rs in a high-quality bug report are an enumerated list of GUI interactions that a user performs on the app sequentially to reproduce the bug. Each step should be atomic: it should represent a single GUI action (\eg\ a tap or type) on a particular GUI component (\eg\ button or text field).
An example of high-quality S2R is ``Tap the `More options' button''. 
\rev{We adapt the \textbf{S2R quality model} from prior work~\cite{Chaparro2019} and define four categories: (1) \textbf{Correct Step (CS)} corresponds to a specific GUI interaction in the application; (2) \textbf{Ambiguous Step (AS)} corresponds to multiple interactions on GUI components; (3) \textbf{Extra Step (ES)} is not required to reproduce the bug; and (4) \textbf{Missing Step (MS)} is a GUI interaction required to reproduce the bug, missing in the report.
\looseness=1
}

\begin{figure}[t]
    \centering
    \includegraphics[width=0.85\linewidth]{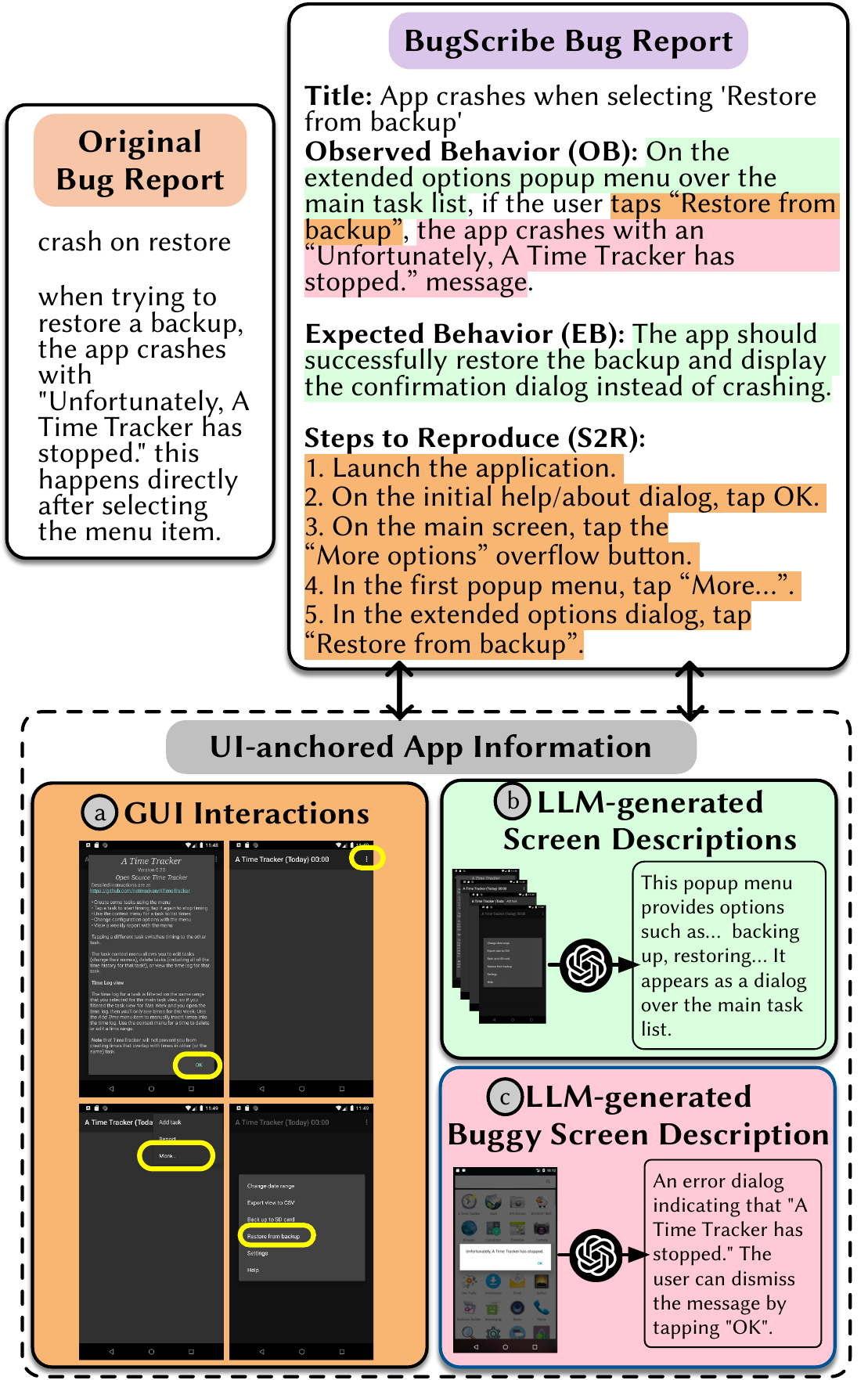}
    \caption{Bug report \#35 of the \rev{ATimeTracker} app~\cite{bugID35}, and the corresponding \tool report. Pieces of the report are linked to: (i) GUI interaction data (orange), (ii) LLM-generated descriptions of bug-related screens (green), and (iii) LLM-generated description(s) of the buggy screen(s) (red).}
    \label{fig:example}
    \vspace{-0.5cm}
\end{figure}

Compared to the original quality model~\cite{Chaparro2019}, we include the Extra Step (ES) category because automated approaches may generate steps that are unnecessary to reproduce a bug, and we discarded the Vocabulary Mismatch (VM) category~\cite{Chaparro2019} because we expect LLMs to avoid this problem. Also, we did not observe this quality issue in our experiments. An example of an \textit{Ambiguous} S2R would be ``\rev{after selecting the menu item}'' (see \cref{fig:example})\rev{,} because \rev{it does not specify which menu and item to select}. When generated S2Rs are compared against ground truth S2Rs, CS, ES, and MS correspond to true positives, false positives, and false negatives, respectively, from which we compute precision, recall, and F1 score.
\looseness=-1

\textbf{OB Quality Model.} A high-quality OB should be a clear, detailed, and self-contained description of the bug and should specify three main information elements: \rev{(1)} the \textit{buggy app behavior} (\ie the observed app misbehavior); (2) a \textit{buggy screen reference} (\ie the screen where the user observed the bug); and (3) the \textit{triggering GUI interaction} (\ie the interaction that triggers the buggy behavior). An example of a high-quality OB is ``The app crashes if I tap the `Restore from backup' option on the pop-up menu of the Main Task screen''. Our model defines \rev{five} quality attributes for each of the OB elements. An \textbf{OB element} can be:
\looseness=-1
\begin{itemize}
    \item \textbf{Correct:} the element corresponds to plausible buggy app behaviors, screens, or GUI interactions.

    \item \textbf{Incomplete:} the element lacks information to fully map it to plausible buggy app behavior, screens, or GUI interactions.

    \item \textbf{Ambiguous:} the element contains ambiguous or generic information for mapping it specifically to plausible buggy app behavior, app screens, or GUI interactions.

    \item \textbf{Incorrect:} the element describes information that does not map \rev{to} the buggy app behavior, screen, or GUI interaction.
    \looseness=-1

    \item \textbf{Missing:} the element is not provided in the OB description.
\end{itemize}

For example, in the OB ``The app crashes if I tap the `Restore from backup' option'',  the \textit{buggy screen reference} is \textit{Missing}, and if it is written as ``The app does not work if I tap the `Restore from backup' option'', it would also contain an \textit{Ambiguous} \textit{buggy app behavior}, and if it was written as  ``The app does not work if I `Restore from backup''', it would also contain an \textit{Incomplete} \textit{triggering GUI interaction}.
\looseness=-1

\textbf{EB Quality Model.} The EB in a high-quality report should be a clear, detailed, and self-contained description of the \textit{intended app behavior}, and it is tied to the OB as it describes the opposite (or an alternative) to the OB's buggy behavior. For example, for the OB ``The app crashes if I tap the `Restore from backup' option on the pop-up menu...'', the EB should be ``The app should successfully restore the backup and display a confirmation dialog''.  We define the same five quality attributes as for the OB elements (\ie correct, incomplete, ambiguous, missing, and incorrect).  
For instance, an EB with an \textit{Ambiguous} \textit{intended behavior} would be ``The app should work''.

\textbf{Motivating Example.}
\Cref{fig:example} shows bug report \#35 submitted for the ATimeTracker app~\cite{bugID35}, which allows users to track the time of their daily activities. This report is unstructured and contains an OB with \textbf{missing EB }(\ie the intended behavior is missing) and \rev{lacks most S2Rs}. The OB describes a correct buggy behavior (\textit{``the app crashes with...''}) but contains an \textbf{incomplete triggering GUI interaction} (\textit{``trying to restore a backup''}). The report contains an \textbf{ambiguous S2R} (\textit{``after selecting the menu item''}). 
\looseness=-1

 \Cref{fig:example} also shows the bug report that our approach, \tool, generates. Unlike the original bug report, this report is structured and provides complete, correct, and detailed OB, EB, and S2R descriptions. For example, the OB includes \textit{correct} descriptions of the buggy behavior (\textit{``...the app crashes with...''}), the \rev{buggy screen reference} (\textit{``On the extended options popup menu...''}), and the triggering GUI interaction (\textit{``... the user taps `Restore from backup'...''}). The S2Rs describe a complete reproduction scenario with atomic steps that correspond to specific GUI interactions (see \cref{fig:example}.a).

%% file: 3_approach.tex
\section{\ap: Generating Enhanced Bug Reports}
\label{sec:approach}

We introduce \ap, an LLM-based automated approach that receives an existing bug report and produces a high-quality report that more effectively describes the equivalent defect from the original report. \ap leverages dynamic app analysis, LLM-based language reasoning/generation, and uses app-related information as context for the LLM to generate bug reports with clear, detailed, and accurate OB, EB, and S2Rs. Based on these components, \textit{\ap also generates a one-line bug summary as bug report title and adds environment details described in the original bug report,} including mobile device information and Android version.

In its current version, \ap uses \gpt~\cite{openai2026gpt54}.
In \ap's evaluation, described in \Cref{sec:evaluation}, we instantiate \ap with \claude~\cite{anthropic2025claudeopus46}, another state-of-the-art LLM, to assess whether using different frontier models impacts the effectiveness of \ap.
\looseness=-1

\begin{figure*}[t]
    \centering \includegraphics[width=0.97\linewidth]{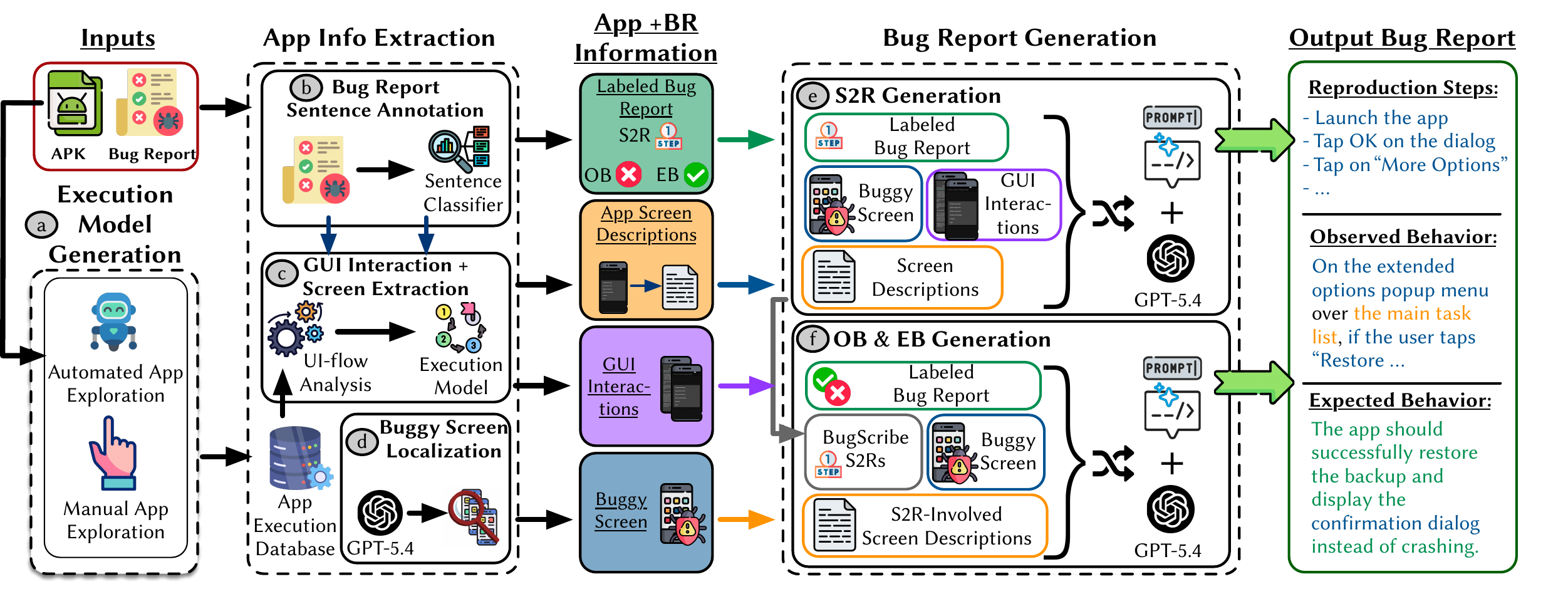}
    \caption{\ap's architecture}
    \label{fig:approach}
\end{figure*}

\ap performs three main phases (see \Cref{fig:approach}):
\begin{enumerate}
    \item \textbf{App Execution Model Generation}: it constructs a graph-based execution model of the app using both automated and manual app exploration (\Cref{sec:execution-model}).

    \item \textbf{Contextual Information Extraction}: it extracts and formats app-related information (\eg\ GUI interactions) used as context for the LLM (\Cref{sec:context-extraction}).

    \item \textbf{Bug Report Generation}: it generates high-quality bug reports, including a title, OB, EB, and S2Rs, leveraging the app-specific information and LLM-based reasoning via zero-shot, task-decomposition prompting (\Cref{sec:br-generation}). 
    \looseness=-1
\end{enumerate}

\subsection{App Execution Model Generation}
\label{sec:execution-model}

In the first phase, \ap constructs a graph-based execution model that captures GUI interactions and screen metadata, including UI hierarchies and events (\Cref{fig:approach}.a). It then extracts and formats this information as context for the LLM to generate high-quality bug reports (\Cref{sec:context-extraction,sec:br-generation}). 
\looseness=-1

\ap represents the execution model as a directed graph, 
where nodes represent unique GUI screens and edges represent user interactions (\eg taps). Screens are distinguished by their UI component hierarchies. Each interaction is represented as $(v_x,v_y,e,c)$, where $c$ is the interacted GUI component (\eg a button) on screen $v_x$, $e$ is the performed action, and $v_y$ is the resulting screen. Each edge also stores metadata such as the component type, ID, label, and description.

\ap constructs the execution model from GUI interaction traces collected through automated exploration and manual app usage \rev{following prior work~\cite{Chaparro2019,saha2024toward}}. For automated exploration, it uses a modified version of \CrashScope~\cite{Moran2016,Moran2017}; manual traces complement workflows that automated exploration cannot cover. 
\looseness=-1

\rev{Execution model generation is a one-time activity for a given app version and does not depend on a specific data collection strategy. Automated exploration requires no human effort beyond running the exploration tool. Manual traces do not need to be collected specifically for \ap: they can come from existing testing activities, such as manual testing, which most mobile app testing still relies on
~\cite{linares2017developers}.
Maintaining and extending the model is also inexpensive: new traces are merged into the existing graph automatically, with duplicate screens identified by their UI hierarchies~\cite{Chaparro2019}. Thus, the model grows as the app is tested and used, and a single model serves all bug reports of the same app version. 

\ap does not require a complete execution model of the app, nor a model that contains the exact reproduction path of a bug. It requires the model to cover the app feature involved in the bug, \ie the screens and interactions related to the buggy screen. 
If the model has no information about the buggy feature, \ap fails to connect the report to the app execution, and its output is limited to the information in the original report. The execution models used in our study cover 52.6\% of the app activities, on average.}

\subsection{Contextual Information Extraction}
\label{sec:context-extraction}

\ap annotates the user-submitted bug report to specify three components (OB, EB, and S2Rs) and extracts three app information elements (app screen information, GUI interactions, and buggy screen) as context for the LLM.
\looseness=-1

\subsubsection{OB/EB/S2R Sentence Annotation}
\ap identifies the \textbf{sentences that describe the OB, EB, and S2Rs} in the input bug report (\Cref{fig:approach}.b). 
\ap decomposes the report into individual sentences and uses \gpt with zero-shot, rule-based prompting to classify each sentence as OB, EB, S2R, or other. The labeled report is then passed to the next phase. 
\looseness=-1

Similar to prior work~\cite{mahmud2025combining}, we designed a prompt to classify OB, EB, and S2R sentences.
The template (found in our replication package~\cite{package}) includes the OB/EB/S2R definitions with detailed guidelines to classify all sentences in the bug report.
\looseness-1

\subsubsection{App Screen Description Generation}

\ap uses \textbf{app screen descriptions} generated by the LLM as additional app context. These descriptions summarize each screen's functionality, layout, GUI components, and available actions. \ap generates one description for every identified screen in the execution model using zero-shot prompting on textual screen metadata (\eg layout and UI component information). 
An example of an app screen description is: ``This popup menu provides options such as changing the date range, exporting to CSV, ... It appears as a dialog over the main task list.''
This context can help the LLM identify the interactions mapped to a reproduction scenario, generate atomic S2Rs, and localize the suspected buggy screen (\Cref{subsec:screen_locatization}). The prompt template is available in our replication package~\cite{package}.
\looseness=-1

\subsubsection{GUI Interaction Extraction}
\label{sec:gui_interactions}

\textbf{GUI interactions} are the building blocks for \tool to generate atomic S2Rs because the interactions required to reproduce a bug are a subset of all possible app interactions. \tool uses the LLM to map the S2Rs in the original report to a navigable interaction sequence in the execution model that leads to the buggy behavior. To achieve this, \ap extracts all GUI interactions from the execution \rev{model} \rev{(\cref{fig:approach}.c}) and provides them to the LLM as context. Each interaction is represented as a tuple containing (1) the action (\eg tap), (2) the target GUI element (\eg ``OK'' button), (3) the source screen ID, and (4) the target screen ID. The LLM then synthesizes these interactions into atomic S2Rs. When S2Rs are missing (\ie there are gaps in paths in the execution model), it infers plausible steps from the OB, EB, other bug report information, and the execution model. 
\looseness=-1

\subsubsection{Buggy Screen Localization} 
\label{subsec:screen_locatization}

The \textbf{buggy screen} is the screen where the faulty behavior occurs. It serves as the ``stopping'' point for S2R generation (\ie the final screen in the reproduction scenario) and provides context for OB/EB generation.
\looseness=-1

\ap identifies the buggy screen by reasoning over the reported bug and the generated screen descriptions (\Cref{fig:approach}.\rev{d}). We formulate this task as a screen retrieval problem~\cite{saha2024toward}, where the LLM ranks all screens in the execution model using zero-shot prompting. \ap selects the top-ranked screen as the buggy screen for subsequent bug component generation. %
To this end, we use LLM-based reasoning with prompts that include the labeled bug report, screen descriptions, and the associated GUI interactions.

\subsection{Bug Report Generation}
\label{sec:br-generation}

\rev{Both generation steps use zero-shot, task-decomposition prompts. Their structure and included app context were determined with the data-driven methodology (\Cref{sec:context_evaluation}).}
\looseness=-1

\subsubsection{S2R Generation}

To generate an improved bug report, \ap first generates atomic S2R descriptions for the input bug report with \gpt (\Cref{fig:approach}.\rev{e}). 
This step uses a zero-shot, task-decomposition prompt to guide the LLM to map the labeled S2R sentences to GUI interaction sequences and infer plausible reproduction scenarios that lead to the buggy screen.
The prompt instructs the LLM to identify the most likely reproduction scenario given the described bug and context and generate a detailed, enumerated list of reproduction steps.
\looseness=-1

\subsubsection{OB/EB Generation} 

\tool generates the OB and EB for the input report using \gpt (\Cref{fig:approach}.\rev{f}), with a zero-shot, \rev{task-decomposition} prompt to guide the LLM to map the identified OB and EB from the original bug report to the contextual app information, including the suspected buggy screen, the S2Rs generated in the previous pipeline step, and the source and target screen descriptions for each generated S2R. This step also generates a bug report title and additional environmental information, such as the Android version and device.
\looseness=-1

Finally, \ap assembles the complete bug report by appending the generated output into labeled sections in the following order: title, OB, EB, S2Rs, and additional information. \rev{Similar to retrieval-augmented generation~\cite{lewis2020retrieval}, \ap provides external information to the LLM. However, \ap does not retrieve this information based on similarity. Instead, it provides all GUI interactions to the LLM to identify the reproduction path, localizes the buggy screen using LLM-based ranking, and provides different app information to generate each report component (\eg the generated S2Rs and their screens for the OB/EB).}
\looseness-1

%% file: 4_prompt_dev.tex
\section{App-\rev{Specific} Information for \ap}
\label{sec:context_evaluation}

We used a data-driven approach to identify the most effective combinations of app-specific information (\eg\ GUI interactions and buggy screens) for \tool's S2R and OB/EB generation. Using a development set, we evaluated reports generated by \ap against ground truth OB, EB, and S2Rs that we manually created based on our bug report quality model. This section describes the development set, the configurations of app information, the prompt design, and the quality assessment procedure, which we also use in \ap's evaluation on the test set (\Cref{sec:evaluation}).
\looseness=-1

\subsection{Development Dataset}
\label{sec:dev-data}

We collected the 10 bug reports of the development set from prior work on S2R quality assessment~\cite{mahmud2025combining}. The reports span nine Android apps from various domains (finance tracking, file management, multimedia, \etc) and various bug types: crashes (3), cosmetic issues (3), output problems (3), and a navigation bug (1). The app execution models associated with the 10 reports contain 37.2 UI screens (ranging from 12 to 74) and 80.5 GUI interactions (ranging from 19 to 159), on average. The dataset includes OB/EB/S2R sentences, videos that show how to reproduce the bugs, and screenshots of the exercised app screens, but it does not include ground truth OB, EB, and S2R descriptions. Using the procedure described next, we created ground truth S2Rs with \rev{109} atomic steps (\rev{10.9} per bug on average), each corresponding to one specific GUI interaction, and one OB and one EB description per report with the correct information elements.
\looseness=-1

\looseness=-1

\subsection{Ground Truth Creation}
\label{sec:s2r_gt_creation}

We used the same procedure to create ground truth for the development set and for the test set used in \ap's evaluation (\Cref{sec:test_dataset}). For each report, one author read the reported bug, watched the reproduction video, and manually inspected the nodes and edges of the app execution model to identify a minimal reproduction scenario, \ie a path of GUI interactions that maps to the S2Rs in the report. This author described each GUI interaction in the path as an atomic S2R using the interaction metadata (\ie the action and the interacted GUI component), following the format [\textit{action}] [\textit{GUI component}] (\eg ``tap the OK button''). The resulting screen of the last interaction in the path was taken as the buggy screen. Based on this screen and the OB/EB in the original report, the same author wrote the ground truth OB and EB. For OBs, the format was ``On [\textit{buggy screen reference}], if the user [\textit{triggering GUI interaction}], [\textit{buggy app behavior}]'' (\eg ``On the confirmation dialog, if the user clicks the OK button, the app crashes''). For EBs, the format was ``[\textit{subject}] should/shouldn't [\textit{intended app behavior}]'' (\eg ``the app should take me to the previous screen without crashing''). 
\looseness=-1

A second author validated the \rev{ground truth}, noting potential mistakes and disagreements. The two authors resolved all issues in a discussion session and reached consensus on the final ground truth. Agreement was high for both datasets. On the development set, the agreement rate was 91.7\% for S2Rs, with the most common disagreement being the misinterpretation of execution model metadata. On the test set, it was 93.5\% for S2Rs and 89.6\% for OBs/EBs.
\looseness=-1

\subsection{Configurations of App Information Used as LLM Context}
\label{sec:configs}

\subsubsection{Configurations for S2R \rev{Generation}}
\label{sec:s2r-config}

\rev{We} designed four cumulative configurations that combine three app information elements. \ConfigNoInfo{} uses the original report and no app information, which reveals the LLM's ability to generate S2Rs without context. \ConfigInt{} adds all \textit{GUI interactions} of the execution model, \ConfigIntScr{} adds the \textit{screen descriptions}, and \ConfigIntScrBug{} adds the suspected \textit{buggy screen}. GUI interactions are expected to let the LLM identify the required steps from the original report, screen descriptions to help it map interaction sequences to the report, and the buggy screen to mark the stopping point of the reproduction scenario, since it is the final screen where the bug manifests.

\subsubsection{Configurations for OB/EB \rev{Generation}}
\label{sec:ob-eb-config}

\rev{We} designed four cumulative configurations that combine three app information elements. \ConfigNoInfo{} uses no app information and serves as a bare-minimum baseline. \ConfigBuggyScr{} adds the \textit{buggy screen description}, \ConfigStoRBuggyScr{} adds the atomic \textit{S2Rs generated} in the S2R phase, and \ConfigStoRAllScr{} adds the \textit{screen descriptions} of the generated S2Rs. The buggy screen description is expected to provide key information such as the OB's buggy screen reference, since the bug manifests on this screen, and the generated S2Rs and their screen descriptions to help the LLM identify the triggering GUI interaction.

\subsection{Prompt Design}
\label{sec:prompt_design}

We designed two core prompt templates, one for S2R generation and one for OB/EB generation (\Cref{sec:br-generation}). Both are zero-shot, task-decomposition prompts~\cite{khot2022decomposed,dua2022successive} that contain the task description, a description of the inputs (\ie the report and app information used as context), the inputs themselves structured into well-defined sections, detailed instructions to complete the task, and the output format. To evaluate the configurations in \Cref{sec:configs}, we instantiated each template with the inputs of each configuration and specific instructions to handle them, \eg to form a valid and complete path in the execution model that ends at the buggy screen. The wording and structure of all instances are otherwise the same.

For each task, one author designed the template following the OpenAI guidelines~\cite{openai_prompt} and prompt engineering best practices~\cite{white2024chatgpt,santana2025prompting,schulhoff2024prompt}, and refined it iteratively using three development bugs and meta-prompting with ChatGPT~\cite{zhang2023meta}. A second author independently reviewed the templates, evaluated them on the same bugs, and suggested improvements. Both authors discussed the suggestions and finalized the templates.

\looseness=-1

\subsection{Prompt Execution and Quality Assessment}
\label{sec:eval_method_dev}

We executed the eight prompt configurations (four per generation task) using the OpenAI API~\cite{openai_api} with the \gpt model. Since the latest GPT models do not have temperature control, we used the model as is. To assess the consistency of LLM generation, we executed each configuration three times and report the average quality across the runs.
\looseness=-1

We assessed the quality of the generated S2Rs, OBs, and EBs by manually comparing them against the ground truth using our quality model (\Cref{sec:example_q_model}). For S2Rs, we labeled each generated step as correct (CS) or extra (ES) and each ground truth step absent from the report as missing (MS), and computed precision, recall, and F1 score. For OBs/EBs, we labeled each information element (\eg the OB's triggering GUI interaction) with one of the model's five quality categories and counted the elements in each category.

To mitigate bias, two authors independently assigned the labels following detailed criteria and annotation guidelines (see replication package~\cite{package}). Disagreements were resolved through discussion, and unresolved cases were evaluated by a third annotator to reach consensus. We measured inter-annotator reliability with observed agreement, Cohen's $\kappa$~\cite{Cohen}, and Krippendorff's $\alpha$~\cite{krippendorff2018content}, interpreting $\kappa$ and $\alpha$ with the scales of prior work~\cite{viera2005understanding,marzi2024k}. Agreement was high: \devSTRagr\ for S2Rs ($\kappa = \devSTRck$, almost perfect; $\alpha = \devSTRka$, reliable) and \devOBagr\ for OBs/EBs ($\kappa = \devOBck$, moderate; $\alpha = \devOBka$, reliable). We use the same assessment procedure in \ap's evaluation (\Cref{sec:evaluation}).
\looseness=-1

\subsection{Results and Analysis}

\begin{figure}
    \centering
    \includegraphics[width=1\linewidth]{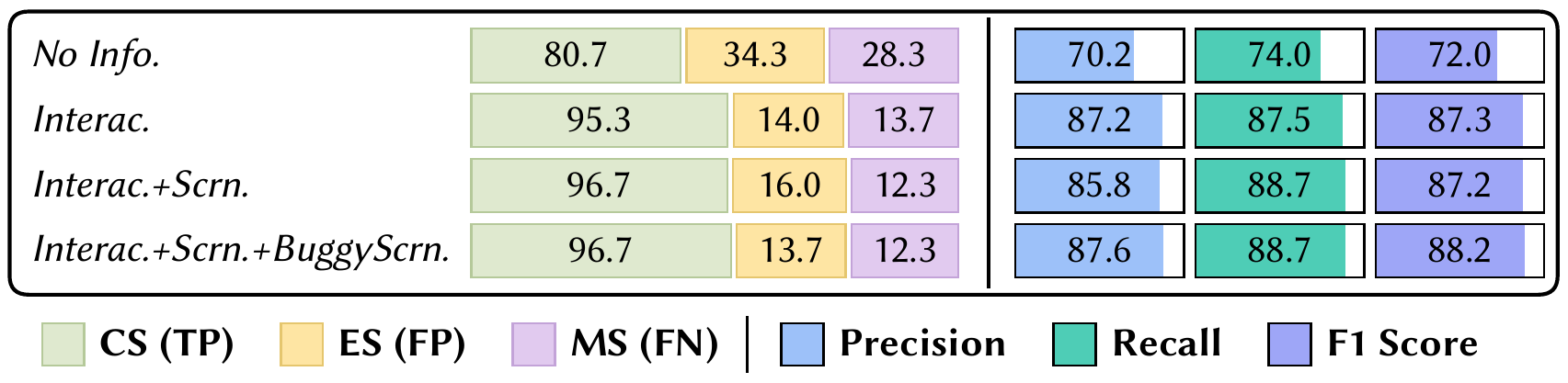}
    \caption{Total \# of correct (CS), extra (ES), and missing S2Rs (MS)\rev{, and precision, recall, and F1} in the development bug reports (avg. over three runs)}
    \label{fig:s2r_results_dev}
\end{figure}

\subsubsection{Quality of Generated S2Rs}
\label{sec:s2r-dev-results}

\Cref{fig:s2r_results_dev} shows the S2R quality of the four configurations, averaged across the three executions. The results are consistent across runs, \eg the standard deviation of the F1 score of \ConfigIntScrBug{} is \rnum{\ISBfoneSTD}. All three context-aware configurations substantially outperform \ConfigNoInfo{}, with F1 increasing from \rnum{\NIfone} to \rev{\rnum{\ISfone}}\rev{--}\rnum{\ISBfone}. GUI interactions alone yield most of this gain (F1 \rnum{\INTfone} vs. \rnum{\NIfone}), which indicates that this context is essential to identify the reproduction path. \rev{Adding the screen descriptions reduces missing steps (\rnum{\ISms} vs. \rnum{\INTms}), and adding the buggy screen reduces extra steps (\rnum{\ISBes} vs. \rnum{\ISes})}, and \ConfigIntScrBug{} achieves the best performance (F1 \rnum{\ISBfone}, precision \rnum{\ISBprec}, recall \rnum{\ISBrec}). 
\looseness=-1

\subsubsection{Quality of Generated OBs/EBs}
\label{sec:ob-eb-dev-results}

All configurations generate OBs/EBs with similar elements across the three runs. Out of 40 OB/EB elements (4 element types $\times$ 10 bug reports), the number of correct elements ranges from \rnum{\SumNCor} with \ConfigNoInfo{} to \rnum{\SumSSCor} with \ConfigStoRAllScr{}, and no configuration misses any element. The gain comes mainly from the buggy screen, which all context-aware configurations share (\rnum{\SumBCor}\rev{--}\rnum{\SumSSCor} correct elements vs. \rnum{\SumNCor}). \ConfigStoRAllScr{}, which adds the generated S2Rs and their screen descriptions, performs best, with the most correct elements and the fewest incomplete ones (\rnum{\SumSSInc}). The results for each element type, the counts of incomplete, ambiguous, and incorrect elements, and the results of all executions follow the same trends and are available in our replication package~\cite{package}. Based on these results, \ap uses \ConfigIntScrBug{} for S2R generation and \ConfigStoRAllScr{} for OB/EB generation, which indicates that each report component requires different app information\rev{.}
\looseness=-1

\looseness=-1

%% file: 5_evaluation.tex
\section{Empirical Evaluation}
\label{sec:evaluation}

We assess the quality of the generated S2Rs, OBs, and EBs compared to baseline approaches on a larger dataset of \TotalTestBugs\ bug reports\rev{, and the usefulness of \ap's reports for automated bug reproduction}.
\looseness=-1
Specifically, we answer the following research questions (RQs):
\begin{enumerate}[label=\textbullet~\textbf{RQ$_{\arabic*}$:}, ref=RQ$_{\arabic*}$, labelindent=0cm, leftmargin=*]
    \item \label{rq:s2r} What is the quality of \ap's S2Rs compared to baseline and original reports?
    \item \label{rq:obeb} What is the quality of \ap's OBs/EBs compared to baseline and original reports?
    \rev{\item \label{rq:reproduction} How do \tool-generated bug reports help an LLM-based agent reproduce Android bugs?}
\end{enumerate}

\subsection{Test Dataset}
\label{sec:test_dataset}

We evaluated \ap with a dataset of \TotalTestBugs{} bug reports (\ie a \textit{test set}, different from the development set), which enables comparison with baseline techniques (\Cref{subsec:baselines}). We collected these bug reports via stratified random sampling across various bug types from 96 reports used in prior work~\cite{saha2024toward,Chaparro2019}. These  \TotalTestBugs{} reports span \TotalUniqueApps{} Android applications of various domains (web browsing, WiFi diagnosis, finance tracking, \etc) and cover different bug types: output issues (\TotalOutputBugs{}), cosmetic problems (\TotalCosmeticBugs{}), crashes (\TotalCrashBugs{}), and navigation bugs (\TotalNavigationBugs{}). Corresponding app execution models have an average of \tstAvgScrn{} GUI screens (ranging from \tstMinScrn{} to \tstMaxScrn{}; median: \tstMedScrn{}) and \tstAvgInter{} GUI interactions (ranging from \tstMinInter{} to \tstMaxInter{}; median: \tstMedInter{}).
\looseness=-1

As these bug reports lack the ground truth needed to evaluate \ap, we constructed ground truth OB, EB, and S2Rs for each report with the procedure described in \Cref{sec:s2r_gt_creation}. We also assessed the OB/EB information elements and S2Rs of the original reports with our quality model to measure the improvements achieved by \ap.
\looseness=-1

In summary, the \TotalTestBugs{} bug reports include \TotalStRTestSet{} atomic S2Rs (\AVGStRTestSet{} steps per bug), each representing one specific app interaction, \eg taps, long-taps, \etc They also include the high-quality OB/EB ground truth with the correct information elements.
\looseness=-1

\subsection{Baseline \rev{Techniques}}
\label{subsec:baselines}

We implemented \ap using two state-of-the-art LLMs, \ie\ \gpt~\cite{openai2026gpt54} and \claude{}~\cite{anthropic2025claudeopus46}, resulting in two variants: \bsgpt and \bscld. 
\looseness=-1

We evaluated both variants against a recent bug report improvement technique, \uts~\cite{acharya2025can}. We also implemented two baselines (\gptnoinfo and \cldnoinfo) which leverage the same LLMs without using any app-specific information. 
\looseness=-1

\subsubsection{\gptnoinfo and \cldnoinfo}
These baselines generate bug reports following a similar generation pipeline as \ap, including the same prompt templates for S2R and OB/EB generation. However, these prompt templates do not include the placeholders for app-specific information and the specific instructions to handle it, \ie\ the \ConfigNoInfo{} configuration (see \cref{sec:configs} for details). We designed \rev{these baselines} to understand how \gpt and \claude perform for bug report generation without app-specific information.
\looseness=-1

\subsubsection{\uts} 
is an LLM-based approach to transform unstructured bug reports into structured bug reports with explicit OB, EB, and S2Rs~\cite{acharya2025can}. 
The original approach uses \rev{\gptfouro} with three-shot prompting to generate structured bug reports for the given unstructured bug report.
To enable a fair comparison, we replaced its \rev{\gptfouro} model with \gpt. We implemented this baseline based on its companion replication package~\cite{u2sbr_package}.
\looseness=-1

\subsection{Methodology and Metrics}
\label{sec:test_method}

We ran \ap and all baseline techniques on the \TotalTestBugs{} bug reports in the test set and generated high-quality bug reports for the given original bug reports. 
Given that our development experiments found near-perfect consistency in LLM responses under the same inputs across three executions, we executed all the approaches only once in this evaluation.
For \rev{\ref{rq:s2r}}, we manually compared the generated S2Rs to the ground truth S2Rs and computed precision, recall, and F1 score following the procedure in \Cref{sec:eval_method_dev}. For \rev{\ref{rq:obeb}}, we extracted the four information elements from the generated OBs and EBs, compared them with the ground truth, and counted the elements in each quality category (\Cref{sec:eval_method_dev}). \rev{We tested statistical significance with two-sided Wilcoxon signed-rank tests~\cite{wilcoxon1945} on per-report S2R precision, recall, and F1, and with exact McNemar tests~\cite{mcnemar1947} on each OB/EB element, using the Holm correction~\cite{holm1979simple} for multiple comparisons.}
The agreement between the two evaluators was 97.3\% for S2Rs ($\kappa = 0.92$, almost perfect; $\alpha = 0.99$, reliable) and 92.9\% for OBs/EBs ($\kappa = 0.81$, \rev{almost perfect}; $\alpha = 0.97$, reliable). Our replication package includes the agreement results for each approach~\cite{package}.
\looseness=-1

\textbf{Comparison Protocol.} As \gptnoinfo, \cldnoinfo, and \uts do not use application information and are not designed to generate atomic S2Rs, they can produce compound S2Rs (\ie\ multiple steps in one S2R sentence), contain different wording than the application vocabulary, or implicitly mention multiple steps. 
To ensure a fair comparison, we adopted the following strategies. If there are \textit{m} atomic steps in the generated bug report, and among them \textit{n} are present in the ground truth as individual steps, we counted \textit{n} correct steps. Moreover, we considered generated steps with different wording or presentation but similar meaning to the ground truth step as \textit{correct}. If multiple steps are implicitly mentioned, we counted them individually. We also used these strategies to compare the S2R quality of the original bug reports against the ground truth, which is required to compare the reports generated by \ap and the baselines with the original reports using our evaluation metrics.
A complete comparison protocol to ensure a fair comparison is found in our replication package~\cite{package}.
\looseness=-1

\rev{\textbf{Bug Reproduction Setup.} For \ref{rq:reproduction}, we used 40 of the \TotalTestBugs{} test set bugs, excluding eight that cannot be reproduced in our setup (\eg four require rotating the device, which the agent does not support). For each bug, we ran five report versions: the original report, the reports generated by \uts, \gptnoinfo, and \bsgpt, and the ground truth report (\ie our manually constructed OB, EB, and S2Rs) as an upper bound.
\looseness=-1

\textbf{Agent Execution.} Also for \ref{rq:reproduction}, we used \mobilerun~\cite{mobilerun}, a widely adopted open-source framework that lets LLM agents perform tasks on mobile devices from natural language instructions.
Its developers report that \mobilerun completes all 116 tasks of the AndroidWorld benchmark for Android agents~\cite{rawles2025androidworld,mobilerun_benchmark}. At each step, \mobilerun captures the UI hierarchy and a screenshot of the current screen via a companion app on the device, and sends them with the task and the previous actions to an LLM. \mobilerun executes the GUI action selected by the LLM (\eg tap, type, or swipe) until the LLM ends the run or 100 steps are reached. 
We selected \mobilerun because its agent interacts with the app only through GUI actions %
and cannot use other means (\eg shell commands) to bypass the reported steps. We did not use bug reproduction techniques (\eg\ \recdroid~\cite{Zhao2019} or \adbgpt~\cite{Feng2024}) because they process the reports with their own NLP or prompting techniques before replaying the bug, which would mix the quality of the reports with the effectiveness of these techniques.
\looseness=-1

We used \mobilerun's direct (single-agent) mode with \gpt, the same LLM used in \tool. Each run started from the same clean emulator snapshot (Pixel 6 with Android 9, or Android 11 for apps that require it), with the app freshly installed and launched. The agent's prompt included the bug report and asked the agent to execute the report's steps in order. Since original and baseline reports could miss steps, we allowed extra actions when needed to carry out a step (\eg passing a welcome screen). We executed each report version three times, resulting in 600 runs (40 bugs $\times$ 5 versions $\times$ 3 runs).
\looseness=-1

\textbf{Judges and Metrics.} %
We used a committee of three multimodal LLMs from different providers as judges: \rev{\opusjudge} via \cldc~\cite{claude_code}, \rev{\gptjudge} via \codex~\cite{codex_cli}, and \rev{\qwenjudge}~\cite{qwen36_openrouter}. For each run, the judges received the ground truth OB and EB, the steps of the report given to the agent (for original reports, the labeled S2R sentences), and the run's execution trace: the screenshot and UI text of each screen, the agent's actions, and the app log. The judges did not know the report version and did not see the agent's own assessment of the run. Each judge decided (1) whether the run reproduced the bug, \ie the app showed the ground truth OB rather than the EB, and (2) whether the agent performed each report step. We used the majority vote as the final label, and measured the \textit{reproduction rate} (\% of bugs reproduced in a run, averaged over three runs) and the \textit{step execution rate} (\% of report steps performed across all runs). 
\looseness=-1

All three judges agreed on 91.0\% of the reproduction decisions ($\alpha = 0.88$, reliable agreement~\cite{marzi2024k}; pairwise $\kappa$ from 0.84 to 0.90, almost perfect agreement~\cite{viera2005understanding}) and on 93.7\% of the step decisions ($\alpha = 0.83$). To validate the committee, one author manually labeled a stratified random sample of 235 runs (95\% confidence level, 5\% margin of error), covering every bug and report version. The committee matched the author's labels for 94.7\% of the reproduction decisions ($\kappa = 0.89$) and 95.8\% of the steps ($\kappa = 0.84$).}
\looseness=-1

\subsection{Results and Analysis}
\label{sec:results}

\subsubsection{\texorpdfstring{\ref{rq:s2r}}{RQ1} -- Quality of S2R Generation}

\begin{figure}
    \centering
    \includegraphics[width=1\linewidth]{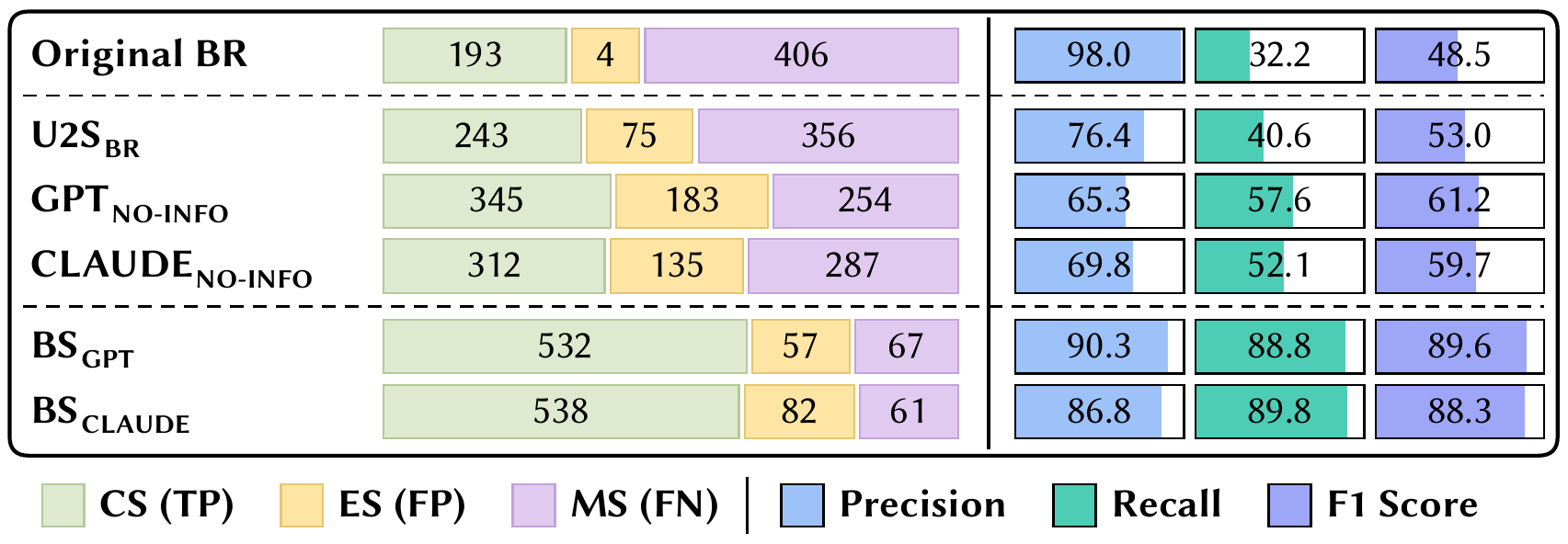}
    \caption{Total \# of correct (CS), extra (ES), and missing S2Rs (MS)\rev{, and precision, recall, and F1} in the test set bug reports}
    \label{fig:s2r_results_test}
\end{figure}

\Cref{fig:s2r_results_test} presents the quality results of the S2Rs in the original bug reports and the S2Rs generated by \bsgpt, \bscld, and the baselines. 
\looseness=-1

\textbf{\ap vs. Baselines and Original Reports.} \rev{The figure} shows that \ap (both \bsgpt and \bscld) clearly outperforms all baselines and the original bug reports across all metrics\rev{, except for the precision and the number of extra steps of the original reports}\rev{, and \bscld's extra steps compared to \uts (82 vs. 75)}. In particular, it generates more correct steps (\rnum{\BSGPTcs}, \rnum{\BSClaudecs} vs. \rnum{\OrigBRcs}, \rnum{\UtwoScs}, \rnum{\GPTNIcs}, \rnum{\ClaudeNIcs}) while reducing both extra and missing steps (\eg \rnum{\BSGPTes}, \rnum{\BSGPTms} vs. \rnum{\UtwoSes}, \rnum{\UtwoSms} by \uts). This results in considerable gains in F1 (\rnum{\BSGPTfone}, \rnum{\BSClaudefone} vs. \rnum{\OrigBRfone}\rev{--}\rev{\rnum{\GPTNIfone}}), meaning that \ap produces higher-quality S2Rs than all alternatives. \rev{\ap's improvements in recall and F1 over the original reports and all baselines, and in precision over all baselines, are statistically significant ($p<0.01$).}
\looseness=-1

These improvements highlight the importance of app context, which enables the LLM to better map S2Rs to reproduction paths from the app execution model. While baselines without such context (\gptnoinfo, \cldnoinfo, and \uts) still improve over the original bug reports (\eg F1 \rnum{\GPTNIfone}, \rnum{\ClaudeNIfone}, \rnum{\UtwoSfone} vs. \rnum{\OrigBRfone}), they remain far behind \ap. 
Across LLMs, \bsgpt and \bscld achieve similar overall performance, with a slight trade-off: \bscld achieves slightly higher recall (\rnum{\BSClauderec}) and more correct steps (\rnum{\BSClaudecs}), while \bsgpt achieves higher precision (\rnum{\BSGPTprec}) and fewer extra steps (\rnum{\BSGPTes}), indicating more precise S2R generation.
\looseness=-1

The lower performance of the baselines 
stems from the lack of context for S2R generation, as they solely rely on user-submitted reports to generate a possible reproduction path. This leads to either missing steps (\rnum{\UtwoSms} for \uts) or over-generation (183 and 135 extra steps for \gptnoinfo and \cldnoinfo). For instance, in bug report \#84~\cite{bugID84} of the Ultrasonic app, the baselines correctly generate \textit{``enter the server URL''} but introduce unnecessary steps
(\eg \textit{``enter the username''} and \textit{``enter the password''}) and miss required interactions (\eg \textit{``Click OK''} and \textit{\rev{``}Swipe up''} before \textit{``Click Test Connection''}), whereas \bsgpt and \bscld avoid such errors by leveraging app-specific information.
\looseness=-1

\looseness=-1

\textbf{Analysis of Failed Cases.} Despite substantial improvements, \bsgpt and \bscld still miss 67 and 61 steps and produce 57 and 82 extra steps, respectively. Our qualitative analysis of these cases reveals three main causes: (1) multiple feasible paths in the app execution model; (2) very low-quality user reports with incomplete or ambiguous steps; and (3) inaccuracies in buggy screen localization. For example, bug report \#1402~\cite{bugID1402} (Phimp.me app) contains only one vague step (\textit{``Try with different images in share activity''}), leading \bsgpt and \bscld to generate only 6 and 8 correct steps (vs. 13 ground truth steps). With insufficient information, \ap tends to produce a shorter but insufficient path. Addressing this limitation may require combining LLM reasoning with path-finding techniques and human-in-the-loop approaches~\cite{song2022toward} to better identify correct interaction paths associated with the bug.
\looseness=-1

\subsubsection{\texorpdfstring{\ref{rq:obeb}}{RQ2}: Quality of OB/EB Generation}

\begin{figure}
    \centering
    \includegraphics[width=1\linewidth]{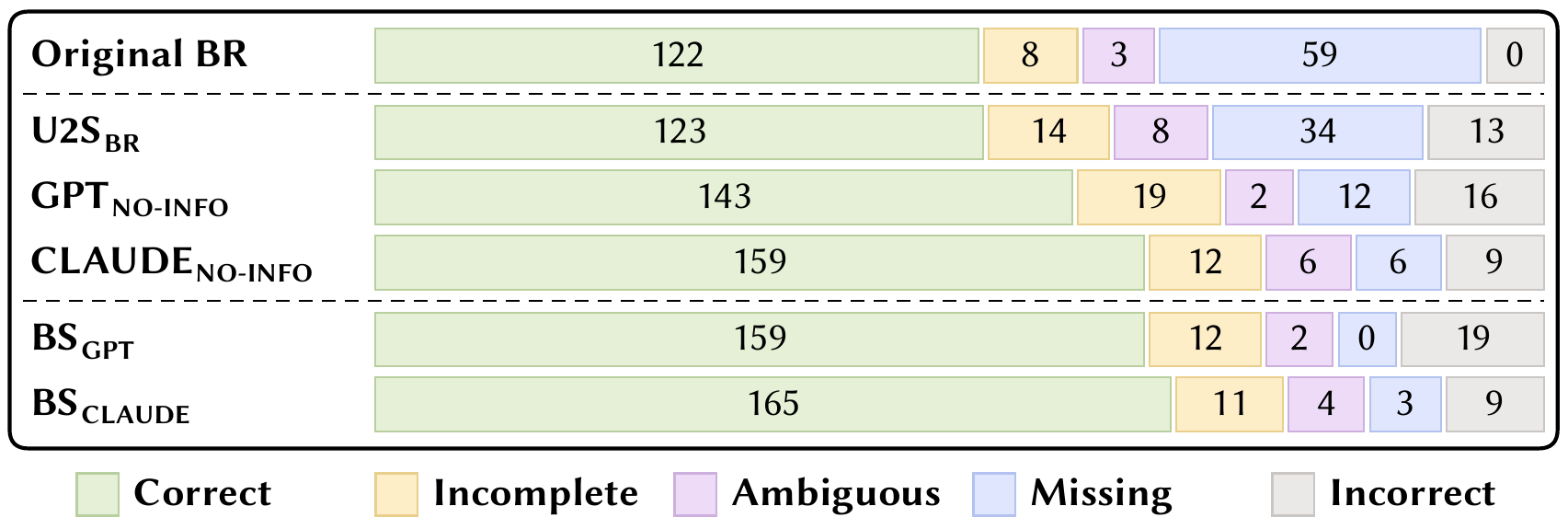}
    \caption{Total \# of OB/EB elements in the \rev{\TotalTestBugs{}} test set
reports}
    \label{fig:ob_eb_results_test}
\end{figure}

\Cref{fig:ob_eb_results_test} shows the total number of OB/EB information elements in the original bug reports and the reports generated using different approaches \rev{(results per element are in our replication package~\cite{package})}.

\textbf{\ap vs. Baselines and Original Reports.} The results vary \rev{across} the individual OB/EB elements. 
(As \uts was not designed to include triggering GUI interaction and buggy screen reference, we abstain from comparing \ap with \uts for these two elements.)
For \textit{Buggy Behavior (OB)}, all approaches perform similarly (\eg 46 correct elements, out of 48), indicating limited impact of app-specific context on generating this element. This can be explained by the fact that the buggy behavior is typically present in original reports \rev{(47 out of 48 reports)}. 
In contrast, for the OB's \textit{Triggering GUI Interaction}, \ap performs better, especially \bsgpt, generating more correct elements (\rnum{\TstTGIbCor}) than the baselines and reducing missing or incomplete cases. For the OB's \textit{Buggy Screen Reference}, results are mixed, but \bscld achieves the highest performance  (\rnum{\TstTSRCCCOr} correct, \rnum{\TstTSRCCCmis} missing elements), while \bsgpt still improves over original reports (\rev{\rnum{\TstTSRBcor}} vs. \rnum{\TstTSROrigCor}).  For \textit{Intended Behavior (EB)}, \ap performs comparably to the baselines, suggesting that context is less important, as intended behavior can often be inferred from the buggy behavior, which appears in \rev{47} of the original reports.
\looseness=-1

\ap improves over the original bug reports, increasing the number of correct elements (\eg\ \rev{159}\rev{--}165 vs. 122) and reducing the number of missing ones (0\rev{--}3 vs. 59), with slight increases in incomplete \rev{(11--12 vs. 8) and more incorrect cases (9--19 vs. 0)}. Compared to the baselines, \bscld achieves the best overall balance\rev{, with the most correct elements (165) and, with \cldnoinfo, the fewest incorrect ones among the generated reports (9)}, while \bsgpt also performs strongly but with \rev{more incorrect elements (19 vs. 9)}. Notably, \cldnoinfo performs \rev{comparably} (with 159 correct elements\rev{, the same as \bsgpt}), suggesting that while context is beneficial, the choice of LLM also influences performance.
\looseness=-1

\rev{Overall, app execution information mainly improves the OB elements that depend on the app execution, \ie the triggering GUI interaction (\eg 42 vs. 28 correct elements for \bsgpt vs. the original reports) and the buggy screen reference (\eg 35 vs. 12 for \bscld). These improvements over the original reports are statistically significant ($p<0.05$), except for the triggering GUI interaction of \bscld.}
\looseness=-1

\textbf{Analysis of Failed Cases.} 
Manual inspection of the incomplete and incorrect triggering GUI interactions generated by \bsgpt/\bscld reveals that extra or unnecessary reproduction steps in the original bug report can misguide \ap. For example, bug report \#1066~\cite{bugID1066} (Focus-android app) contains the line: \textit{``Repeat moving cursor and clicking system status bar...''}, where the user attempted to circumvent the bug. This operation is not a part of the reproduction steps; however, \ap considers this as the triggering GUI interaction, resulting in an incomplete/incorrect element. 
\looseness=-1

We analyzed the ambiguous buggy screen references generated by \ap and found that, although imprecise, they refer to the correct buggy screens. These ambiguities arise because \ap often uses generic, functionality-based names rather than the exact UI wording. For instance, in bug report \#154 for the \rev{CampFahrplan} app~\cite{bugID154}, the correct buggy screen is \textit{``schedule screen''}, whereas \bscld refers to it as \textit{``event details screen''}. This discrepancy stems from limitations in the app metadata used to build the app execution model, which may omit or differ from the user-visible screen names, prompting \ap to infer names from the UI screen descriptions used as context. To address this, our future work will incorporate visual information from UI screenshots via multimodal models to improve screen naming. Besides ambiguous cases, incorrect buggy screens were primarily generated because the buggy screen localization phase failed to identify the correct buggy screen. Our future work will experiment with the inclusion of multiple buggy screen suggestions, \ie not only the most suspicious screen detected by the localization phase.

\subsubsection{\texorpdfstring{\ref{rq:reproduction}}{RQ3}: Bug Reproduction}
\label{sec:usefulness}

\begin{figure}
    \centering
    \includegraphics[width=1\linewidth]{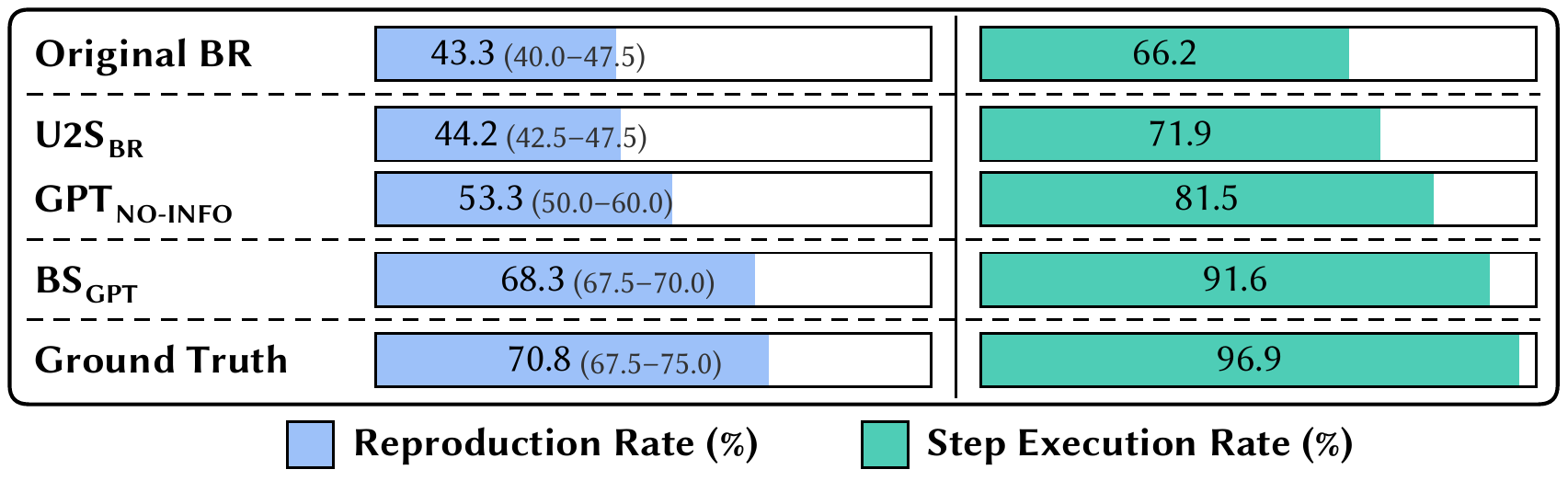}
    \caption{Reproduction rate (avg. and min--max over three runs) and step execution rate for each report version}
    \label{fig:reproduction_results}
\end{figure}

\rev{\Cref{fig:reproduction_results} presents the reproduction and step execution rates achieved by the agent with the original reports, the reports generated by \bsgpt and the baselines, and the ground truth reports.
\looseness=-1

\textbf{\ap vs. Baselines and Original Reports.} With \bsgpt reports, the agent reproduces substantially more bugs than with the original reports and the baseline reports (68.3\% vs. 43.3\%, 44.2\%, and 53.3\%), which represents a 28.1\%\rev{--}57.7\% relative improvement. \rev{The improvements over the original and \uts reports are statistically significant ($p<0.01$).} This rate is close to the one achieved with the ground truth reports (70.8\%). \bsgpt reports also lead to more consistent outcomes: the agent obtains the same outcome in all three runs for 38 of the 40 bugs (95.0\%), compared to 35 (87.5\%), 32 (80.0\%), and 31 (77.5\%) bugs with the original, \uts, and \gptnoinfo reports, respectively. The step execution rate follows the same trend. The agent performs 91.6\% of the steps in \bsgpt reports, compared to 66.2\%--81.5\% for the original and baseline reports and 96.9\% for the ground truth reports.
\looseness=-1

These results are consistent with the S2R quality results (\ref{rq:s2r}). As \bsgpt generates atomic S2Rs that map to valid GUI interactions, the agent can execute each step directly, without guessing how to reach the screen or component the step refers to. In contrast, the original and baseline reports often contain compound or vague steps (\eg \textit{``Navigate to the Advanced Editor''}) or omit required steps, so the agent has to infer the missing interactions and frequently fails to find them. While \gptnoinfo reports improve over the original reports (53.3\% vs. 43.3\%), they remain far behind \bsgpt, as their generated steps are not grounded in the app execution.
\looseness=-1

\textbf{Bugs Reproduced Only with \ap Reports.} Four bugs were reproduced with \bsgpt reports but in none of the runs with the original or baseline reports, and four others were reproduced in all three runs with \bsgpt reports but in at most one run with the other reports. In these cases, \bsgpt recovered required interactions that the original reports omit and the baselines could not infer. For example, bug report \#1028~\cite{bugID1028} (AnkiDroid) only states \textit{``Enter any word into the advanced editor''} and \textit{``Press TRANSLATE''}, without explaining how to open the editor or that the Glosbe dictionary must be selected. With the original and baseline reports, the agent searched for the editor in unrelated screens (\eg Settings), whereas \bsgpt's report includes the full path to the translation. Similarly, the crash in bug report \#8~\cite{bugID8} (Tricky Tripper) requires a traveler in the trip, a precondition that only \bsgpt's report includes.
\looseness=-1
}

\subsubsection{Computational Cost}

We measured the end-to-end cost of \ap. On average, \bsgpt processes 121.8K input and 9.0K output tokens, costing \$0.44 and taking 135 seconds (\rev{$\approx$2.3 min}) per bug report. \bscld processes 154.8K input and 5.6K output tokens, costing \$0.92 and taking 128 seconds. 
Detailed cost data is found in our replication package~\cite{package}.
\looseness=-1

%% file: 7_threats.tex
\section{Threats to Validity}
\label{sec:threats}

\textbf{Construct Validity.} The manual construction of ground truth OB, EB, and S2Rs and the manual assessment of generated reports may introduce subjectivity. To mitigate this, two authors followed the replicable procedures in \Cref{sec:s2r_gt_creation,sec:eval_method_dev}, resolved disagreements through discussion, and obtained high inter-rater agreement. \rev{Since the ground truth S2Rs are atomic and derived from the execution model, our comparison protocol accepts generated steps with different wording or granularity to avoid bias toward \ap (\Cref{sec:test_method}), and the reproduction agent could follow any valid path, even from non-atomic steps.} \rev{In the bug reproduction study, the runs were labeled by LLM judges, which may make mistakes. We used three judges from different providers with majority voting, which led to high agreement, and validated their labels against manual labels, for a \rev{representative sample of runs (95\% confidence, 5\% margin of error)}.}
\looseness=-1

\textbf{Internal Validity.} \ap's performance can vary with the wording and structure of the prompt templates. We designed the templates following best practices~\cite{white2024chatgpt,santana2025prompting,schulhoff2024prompt} and evaluated them under different configurations on the development set. LLM non-determinism may cause results to fluctuate across executions. We ran the development experiments three times and observed consistent results\rev{, and also ran each report version three times in the reproduction study}. Our datasets contain public bug reports and Android projects, so the LLMs may have seen related content during training. All approaches use the same LLMs on the same benchmark, which makes bias toward \ap unlikely. Moreover, \ap guides the LLM to reason over the provided bug report and app execution information rather than over knowledge acquired during training.
\looseness=-1

\looseness=-1

\textbf{External Validity.} Our results may not generalize beyond the bug reports and apps in our datasets. However, the datasets cover various bug types (\eg\ \rev{output issues, cosmetic problems, crashes, and navigation bugs}) and \TotalUniqueApps~Android apps of various domains and sizes, and \ap was developed on a distinct development set, with similar results on both datasets. \rev{The reproduction results are based on one agent, one LLM, and emulated devices, and may differ for other agents, LLMs, or physical devices.
Our study does not measure how developers use the reports (\eg reproduction time), which we leave for future work.} \rev{\ap's effectiveness depends on how well the execution model covers the feature involved in the reported bug. The models in our study include manual traces that exercise the reported features. With models that cover the buggy feature only partially, \ap may generate incomplete S2Rs. 
}
\looseness=-1

%% file: 8_related_work.tex
\section{Related Work}
\label{sec:related_work}

\textbf{Bug Report Quality Assessment and Interactive Bug Reporting.} Prior work has assessed bug report quality through lexical analysis (\eg readability)~\cite{Dit2008,Linstead2009,Zimmermann2010} and the presence of bug information (\eg patches and screenshots)~\cite{Zimmermann2010,Chaparro2019}. The most related work is \EulerC by Chaparro \etal~\cite{Chaparro2019}, which evaluates S2R quality via heuristics that map report keywords to app UI information, and \astro by Mahmud \etal~\cite{mahmud2025combining}, which uses LLMs to assess S2R quality. Imran \etal~\cite{Imran2021} enhance incomplete reports by \rev{selecting} follow-up clarification questions. Interactive bug reporting methods guide reporters during submission. The chatbot \burt~\cite{song2022toward,song2022burt} verifies bug information quality in real time and suggests improvements. \Fusion~\cite{Moran2015} lets reporters build structured reports by selecting actions and GUI components from drop-down lists, and \ebug~\cite{Fazzini:TSE22} extends \Fusion by also suggesting S2Rs. These methods provide feedback on problematic report content but leave corrections to end users, rather than automatically improving reports with app context as \ap does. \rev{\ap builds on the graph-based reporting of \Fusion and \ebug, but it works on existing free-text reports after submission, requires no interaction with the reporter, and generates the OB and EB in addition to the S2Rs.}

\looseness=-1

\looseness=-1

\textbf{Bug Report Quality Improvement.} Prior work on improving bug report quality has focused on detecting missing components and generating structured reports with LLMs. Acharya \etal~\cite{acharya2025can} transform unstructured bug \rev{reports} into structured \rev{ones} with explicit OB, EB, and S2Rs. Bo \etal~\cite{Bo2024} introduced \rev{\chat}, which identifies missing components with a fine-tuned BERT classifier and then generates them with an LLM. Unlike these approaches, \ap formulates bug report enhancement as the problem of connecting user-written reports with app execution. It links report information to relevant execution elements and uses a component-specific grounding strategy to provide the most appropriate context for LLM reasoning, which yields more accurate, complete, and correct OB, EB, and S2Rs, than approaches that rely primarily on the report text. Other work has used LLMs to recommend screenshots~\cite{tan2025imager} and user reviews~\cite{pilone2025automatically} relevant to bug reports, generate bug reports from chat conversations~\cite{shi2022buglistener}, classify report sentences into templates~\cite{zhang2025empirical}, and generate issue report templates~\cite{nikeghbal2024girt}.
\looseness-1

%% file: 9_conclusions.tex
\section{Conclusions}
\label{sec:conclusions}

We showed that automatic bug report enhancement can be achieved by connecting user-written bug reports with application execution. Through a component-specific grounding strategy, \ap generates higher-quality \rev{S2Rs} than the original reports and LLM-based approaches that rely primarily on report text\rev{, and OB/EB that are better than the original reports and at least as good as these approaches}\rev{, with the largest improvements for S2Rs and the execution-dependent OB elements}. \rev{\ap's reports also help an LLM-based agent reproduce more bugs, and more consistently, which illustrates their usefulness for downstream tasks such as bug reproduction.}

\looseness=-1